\documentclass[aps,prl,twocolumn,groupedaddress]{revtex4-1}
\usepackage[latin9]{inputenc}
\usepackage{verbatim}
\usepackage{graphicx}
\usepackage{xargs}[2008/03/08]
\usepackage{times}
\usepackage{amsmath}
\usepackage{geometry}
\usepackage{braket}

\begin{document}

\preprint{}

\title{Measuring the quantum state of a photon pair entangled in frequency and time}

\author{Alex O. C. Davis$^{1}$}
\author{Val\'{e}rian Thiel$^{1}$}
\author{Brian J. Smith{$^{1,2}$}}
\affiliation{{$^{1}$}Clarendon Laboratory, University of Oxford, Parks Road, Oxford, OX1 3PU, UK}
\affiliation{{$^{2}$}Department of Physics and Oregon Center for Optical, Molecular, and Quantum Science, University of Oregon, Eugene, Oregon 97403, USA }


\date{\today}


\maketitle



\textbf{Measuring high-dimensional quantum states of light is an essential capability for all optical quantum technologies including quantum key distribution \cite{tittel:00, nunn:13} and quantum-enhanced sensing \cite{raymer:13} and communications \cite{bennett:99}. To access the advantages of using quantum states, these measurements must be sensitive to the nonclassical correlations that can exist between separate subsystems, such as entanglement \cite{martin:17}. However, until now a flexible approach to determining the full quantum state of systems of multiple time/frequency-entangled photons has been lacking. Here, we present a technique to completely characterise the spectral-temporal wave function of a broadband photon pair using spectral-shearing interferometry. Our method is fully self-referencing and is generalisable across a wide range of wavelengths. To demonstrate, we generate an entangled photon pair with controllable hybrid time-frequency entanglement and perform a full reconstruction. These results allow previously unobservable features of quantum states to be measured and utilised for the first time. We foresee this will unlock new opportunities in high-dimensional quantum technology, enabling the development of novel ways of generating and interacting with quantum systems.}

Quantum correlations between photon pairs can arise in any of the physical degrees of freedom of light, but the time-frequency (TF) space as a basis for encoding quantum information is of paramount interest due to its large information capacity and compatibility with integrated optical platforms. The TF domain is also attractive because there are well-developed technologies to produce photon pairs with high-dimensional TF entanglement \cite{brecht:15}. Recently, strides have been made towards both controllably manipulating \cite{roslund:14,matsuda:16,karpinski:17, wright:17, allgaier:17} and measuring \cite{wasilewski:07, qin:15,davis:17} the TF properties of single photons. For many applications of quantum light, however, it is insufficient to know only the individual states of separate quantum systems: the global wave function of a wider entangled system must be determined through coincident measurements on each of the subsystems. The ability to perform such a measurement has diverse applications across all stages of quantum technology, from the development of entanglement sources to understanding the behaviour of quantum devices such as interfaces, memories, repeaters and gates []. 

Full TF characterisation of single photons is challenging, since common methods of classical pulse characterisation rely on nonlinear effects which are too weak at the single-photon level \cite{trebino:97,walmsley:09}. Such techniques also require spectrally-resolved detection of single photons, which is often prohibitively noisy and inefficient. Other methods use interference with an external mode-matched reference field, but this introduces the need for a stable and tunable optical reference and a faithful \emph{a priori} estimate of the state which severely limits the range of unknown inputs that can be reliably characterised \cite{qin:15, wasilewski:07,polycarpou:12}. These problems have so far obstructed efforts to advance beyond the TF characterisation of a single system to multi-photon systems. Whilst measurements of some of the  properties of the two-photon state, such as joint spectral intensity (JSI), have been possible for some time, these are phase-insensitive and hence cannot reveal the full extent of nonclassical correlations \cite{zielnicki:18}. Narrowband photon pairs have been fully characterised using one another as a reference and using a variable delay followed by time-resolved measurement \cite{chen:15}, although this method is restricted to spectrally narrow biphotons with good spectral and spatial overlap. Recently, we demonstrated a robust, self-referencing approach to single-photon TF wave function characterisation based on an electro-optic spectral shearing interferometer (EOSI) \cite{PRL:18,PRA:18}. Here, we demonstrate how a generalisation of this method may be used to fully characterise the wave function of an entangled multi-photon system. We send one photon from a TF-entangled pair into the EOSI whilst spectrally resolving the other and recording the measurement outcomes in coincidence. By repeating this measurement in the other configuration, all TF information about a pure photon pair can be recovered. We also propose how to generalise this method further to enable TF characterisation of spectrally mixed states. This method remains fully self-referencing and introduces no constraints on the mode overlap between the two photons themselves.

\begin{figure}[t]
	\centering
	\includegraphics[width=.8\linewidth]{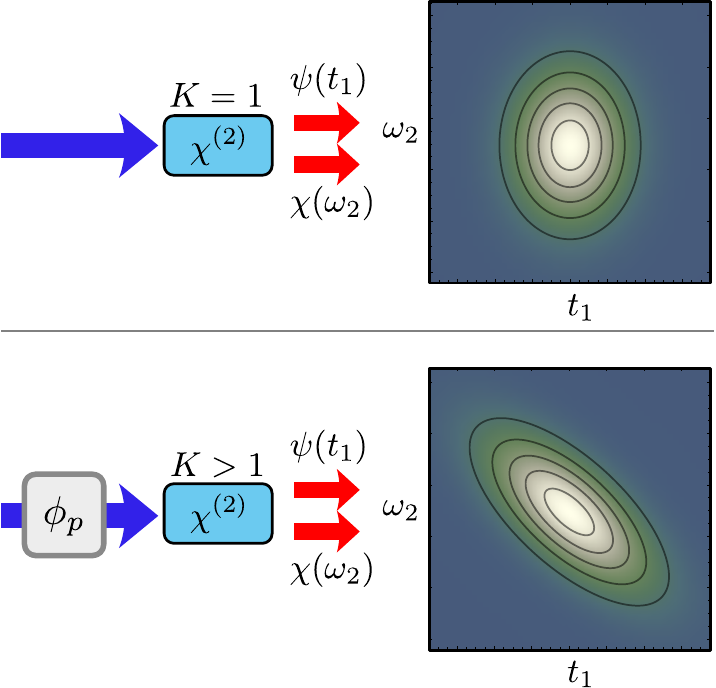}
	\caption{Schematic depicting the effect of pump's spectral phase on entanglement. SPDC in the non-linear medium results in a non-entangled state in both temporal and spectral domain (a). The same process with an additional spectral phase $\phi_p$ on the pump field results not only in a non-trivial entangled state in both domain, but entanglement also appears between the temporal and spectral mode-functions of the two daughter photons.}
	\label{fig:principle}
\end{figure}

\begin{figure*}[t]
	\centering
	\includegraphics[width=\linewidth]{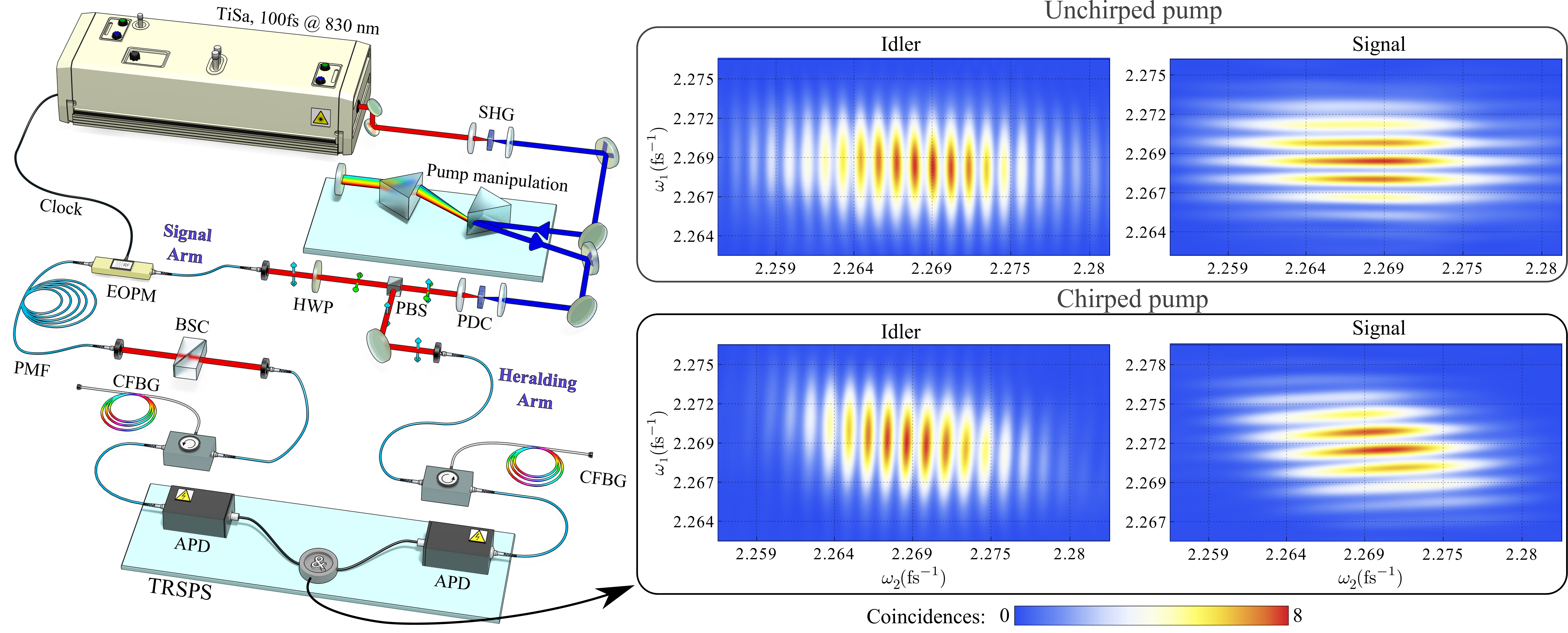}
	\caption{Left: Experimental set-up. 
	SHG- second harmonic generation. TRSPS-time-resolved single-photon spectrometer. BS- beam splitter. PD- photodiode. DM- dichroic mirror. PMF- polarisation-maintaining fibre. SPCM- single photon counting module. CBG- chirped fibre Bragg grating. C- optical circulator. Right: typical lowpass-filtered interferograms for both daughter photons of the SPDC and for an unchirped and chirped pump representing a 6 minutes acquisition.}
	\label{fig:schematic}
\end{figure*}

The pure two-photon TF wave function $\ket{\Psi}$ can be expressed in terms of the complex-valued joint spectral amplitude $f(\omega_1,\omega_2)$ such that
\begin{equation*}
\ket{\Psi} =\int f(\omega_1,\omega_2) a^\dagger(\omega_1)b^\dagger(\omega_2) \textrm{d}\omega_1 \textrm{d}\omega_2\ket{\mbox{vac}}
\end{equation*}
where $a^\dagger(\omega_1)$ and $b^\dagger(\omega_2)$ are bosonic operators that create photons in the signal and idler modes with frequencies $\omega_1$ and $\omega_2$ respectively, and $\ket{\mbox{vac}}$ is the vacuum state. $f(\omega_1,\omega_2)$ is related to the joint temporal amplitude $\tilde{f}(t_1,t_2)$ by a 2-dimensional Fourier transform, and the square modulus of the joint spectral amplitude $|f(\omega_1,\omega_2)|^2$ corresponds to the JSI. 
Since it is a function of two variables, it is possible to perform a Schmidt decomposition on $f(\omega_1,\omega_2)$ \cite{peres:06}:
\begin{equation}
f(\omega_1,\omega_2)=\sum\mathop{}_{\mkern-5mu i} \lambda_i \psi_i(\omega_1)\chi_i(\omega_2)
\label{schmidt_decomposition}
\end{equation}
This represents $f(\omega_1,\omega_2)$ as a sum of straightforward products of normalised functions of just one of the photon's frequencies, $\psi_i(\omega_1)$ and $\chi_i(\omega_2)$, weighted by real coefficients $\lambda_i$. $f(\omega_1,\omega_2)$ is said to be separable if there is only one nonzero coefficient $\lambda_i$, in which case the corresponding $\psi_i(\omega_1)$ and $\chi_i(\omega_2)$ may be interpreted as the wave functions of the two photons and outcomes of local measurements on each photon are not correlated. Otherwise, the two photons are said to be entangled and measurements will exhibit nonclassical correlations. We use the effective Schmidt rank $K$ as our measure of entanglement (see Methods). Values of $K$ greater than unity indicate spectral-temporal entanglement, with $K$ corresponding to the effective number of modes of entanglement.

Our approach to measuring $f(\omega_1,\omega_2)$ relies on electro-optic spectral shearing interferometry \cite{dorrer:03,walmsley:09}. The first photon enters a Mach-Zehnder interferometer where one path involves a constant translation of the spectrum relative to the other (the ``spectral shear", $\Omega$) and the other path imparts a relative delay $\tau$. The spectral shear is applied in an electro-optic modulator (EOM, EOSpacePM-AVe-40-PFU-PFU-83) by the electro-optic effect, the deterministic nature of which ensures unit internal efficiency of the spectral translation. The joint spectral intensity of the first photon upon emerging from one port of the interferometer and the second photon is then

\begin{align*} 
S_\Omega(\omega_1,\omega_2;\Omega)&= \frac{1}{4}\left\{S(\omega_1,\omega_2)+S(\omega_1+\Omega,\omega_2) \right. \\ &+ \left. 2\mbox{Re}\left[f(\omega_1,\omega_2)f^*(\omega_1+\Omega,\omega_2)e^{i\omega_1\tau}\right] \right\}, \nonumber
\end{align*}
where the spectral intensity of the original pulse is given by $S(\omega_1,\omega_2)$. The relative spectral phase can be extracted from the value of the final (spectral interference) term using established data treatment techniques (\cite{walmsley:09}; Methods): 
\begin{equation*}
\phi(\omega_1+\Omega,\omega_2)-\phi(\omega_1,\omega_2)=\mbox{Arg}[f(\omega_1,\omega_2)f^*(\omega_1+\Omega,\omega_2)]
\end{equation*}
Hence, this measures both the joint spectral intensity and the relative phase of $f(\omega_1,\omega_2)$ along lines of constant $\omega_2$. From this it is possible to reconstruct $f(\omega_1,\omega_2)$ up to an additive term that is a function of $\omega_2$ only. This term can be determined by simply swapping the configuration, such that photon 1 is immediately measured spectrally and photon 2 is routed into the interferometer. In the event that the biphoton is not in a pure state, other terms off the main diagonal of the spectral density matrix can be sampled by repeating the measurement for different values of the spectral shear $\Omega$ \cite{PRA:18}.

 To demonstrate, we generate a pair of photons in orthogonal polarisations at 830 nm centre wavelength with full width at half maximum (FWHM) bandwidth of 2.5 nm and 7.5 nm from Type-II spontaneous parametric downconversion (SPDC) in a potassium dihydrogenphosphate (KDP) crystal \cite{mosley:08}. For this demonstration we assume the biphoton is emitted in a pure state. By convention, the narrowband and broadband photons are referred to as the ``signal" and ``idler" respectively. The crystal is pumped by 415 nm femtosecond pulses derived from a frequency-doubled titanium sapphire (Ti:Sapph) laser (see Methods). The photon pair source is designed such that the joint spectral intensity $S(\omega_1,\omega_2)=|f(\omega_1,\omega_2)|^2$ is separable \cite{mosley:08}, and hence the spectral intensities of the two photons are not correlated. However, the complex joint spectral amplitude $f(\omega_1,\omega_2)$ is generally not separable because of correlations in the spectral phases of the single photons. The spectral phase of the pump may be expanded as follows:
 \begin{align}
 \frac{\phi_p}{2} (\omega_1+\omega_2)^2 = \frac{\phi_p}{2}\omega_1^2 + \frac{\phi_p}{2}\omega_2^2 + \phi_p \omega_1 \omega_2
 \end{align}
 
The first two term each depend on the frequency of only one daughter photon so do not affect the Schmidt number. This shows that the dispersion from the pump gets transferred equally on to both photons. The third term, however, cannot be expressed as a separable function and hence generates entanglement, with $K$ increasing monotonically with $\phi_p$. The joint spectrum is hence described by $f(\omega_1,\omega_2) = \psi(\omega_1)\chi(\omega_2)e^{i \phi_p\omega_1\omega_2}$ where the quadratic phase terms in $\omega_1$ and $\omega_2$ have been incorporated into the modes $\psi(\omega_1)$ and $\chi(\omega_2)$ respectively. We can therefore control the entanglement of the biphoton by adjusting the pump dispersion. The nonlocal phase term thus introduced may be interpreted as a time delay on either photon that varies with the frequency of the other, and so it creates correlations whereby the outcomes of spectral measurements on either photon are correlated with the arrival time of the other, and vice versa. Note that any linear optical element between generation and detection can affect the spectral phase of the two modes but may not affect the phase correlations.

It is straightforward to show that the joint temporal distribution has similar expression $\tilde{f}(t_1,t_2) = \tilde\phi(t_1) \tilde\chi(t_2) e^{i \alpha t_1 t_2}$, where $\alpha$ is a coefficient that depends on $\phi_p$ and on the temporal bandwidth of the single photons.
Similarly to the case with joint spectral measurements, if one were to perform a joint time-of-flight measurement on the state, no correlations would be visible since they are carried in the temporal phase. Remarkably, therefore, both joint-spectral and joint-temporal intensity measurements are uncorrelated, and entanglement of the form induced by the nonlocal spectral phase is therefore impossible to establish from such observations alone. By taking the Fourier transform of $f(\omega_1,\omega_2)$ along just one dimension, one obtains the spectral probability distribution of one photon as a function of the arrival time fo the second (fig.\ref{fig:entanglement}). In the present case, with the transform taken over $\omega_1$, this time-frequency amplitude is given by $\tilde{\psi}(t_1-\phi_p \omega_2) \cdot \chi(\omega_2)$.

Once prepared, the two photons are separated at a polarising beam splitter. The idler photon is immediately directed into a time-resolved single-photon spectrometer (TRSPS) \cite{davis:17} where it undergoes a spectral measurement. The signal photon is sent into the EOSI. In our realisation, the two paths of the interferometer are collinear and are distinguished by polarisation \cite{PRA:18,PRL:18} (see Methods). Although monitoring both outputs of the interferometer is possible and would reduce measurement times, it suffices to monitor a single output, where a spectral measurement is recorded by a second TRSPS and logged in coincidence with the outcomes of the measurement for the idler photon. This is then repeated for many copies of the state until significant statistics are obtained. These data may then be represented as a two-dimensional histogram (fig.\ref{fig:schematic}), showing spectral interference fringes. 

\begin{figure*}
	\centering
	\includegraphics[width=\linewidth]{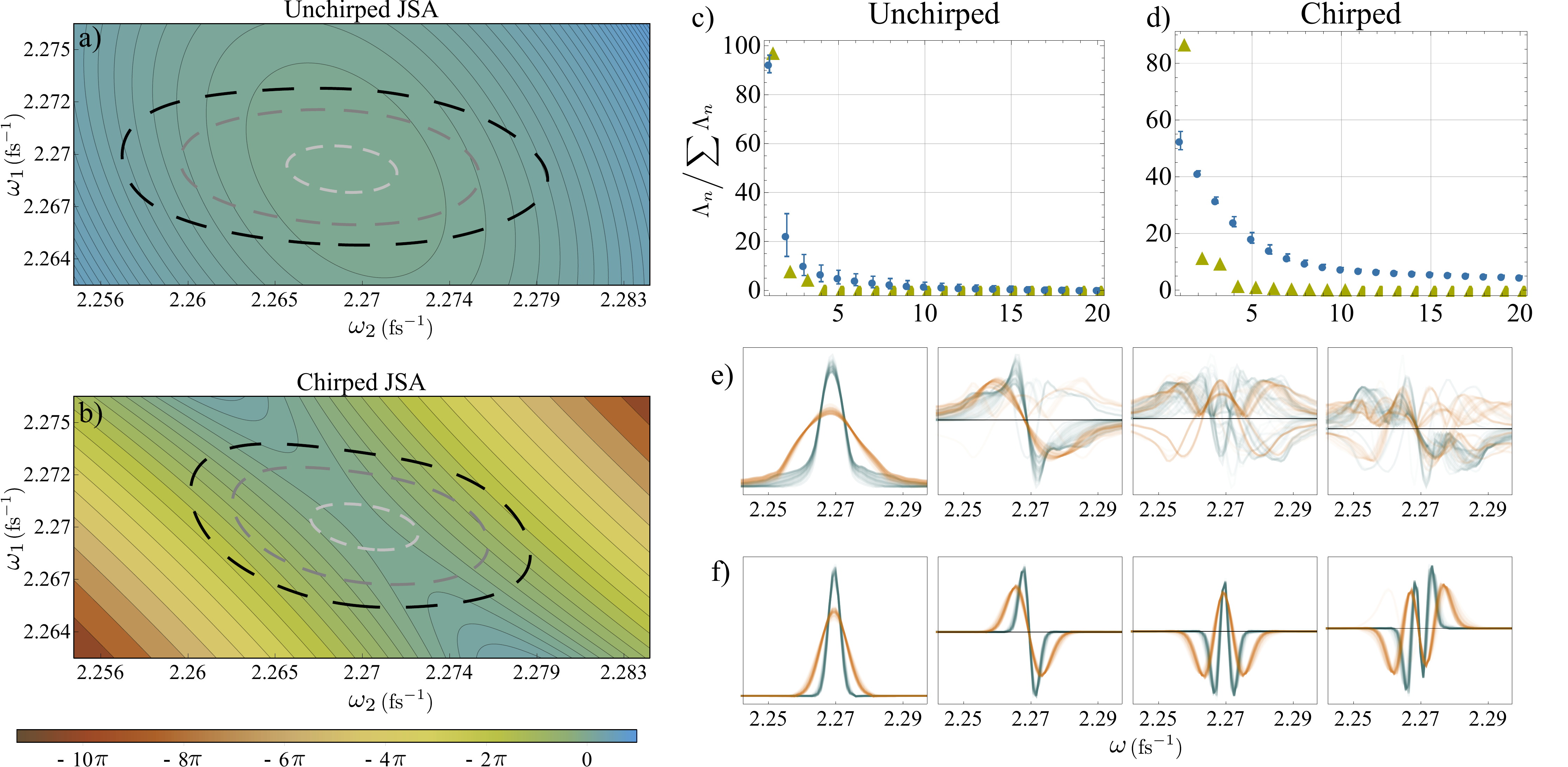}
	\caption{Left: Full phase-sensitive reconstruction of a two-photon joint spectral-temporal wave function for an unchirped (a) and a chirped (b) pump field. Background colour indicates the phase of the complex joint spectral amplitude $f(\omega_1,\omega_2)$ whilst the grayscale contours show lines of constant amplitude $|f(\omega_1,\omega_2)|^2$. Top right (c,d): first 20 singular values of the decomposition of the complex matrices (a) and (b). Diamonds markers represent the diagonalization of the modulus of the matrix, neglecting all phase effects. The error bars represent standard deviation. Bottom right (e,f): overlay of the 4 first singular vectors for an unchirped (e) and a chirped (f) pump.}
	\label{fig:reconstruction}
\end{figure*}

For a chirped pump, the position of these fringes varies as a function of $\omega_2$, giving them a diagonal slant. Hence, the pattern is clearly not the product of its marginals, and so eq.\ref{schmidt_decomposition} implies entanglement in the biphoton. This diagonal slope is due to the nonlocal spectral phase $\phi_p\omega_1\omega_2$. The Fourier transform of the interferogram clearly shows two sidebands that are each offset in the heralding photon direction, pointing towards correlations in phase. Taking the argument of a sideband over both variables yields respectively the spectral phase of the measured photon and the amount of linear correlation. These two measurement allow the retrieval of the same quantity $\phi_p$ although the absolute single photon phase can contain additional information about the optical path. The correlation measurement for the signal photon yields a value of $(-1.61 \pm 0.08) \cdot 10^5$ fs$^2$ while the quadratic spectral phase is estimated to $(-1.42 \pm 0.2) \cdot 10^5$ fs$^2$. The values are negative because the prism compressor adds anomalous dispersion to the pump beam which is passed to the daughter photons. The dispersion of the signal photon is less negative than the correlated phase because of optical path and remaining dispersion within the polarization interferometer which was measured at $\sim 5\cdot 10^3$ fs$^2$.

To complete the full characterisation and compare with the theoretical model, the signal and idler photon were switched. The idler photon was then injected into the EOSI and heralded by the spectrally-resolved signal photon. Fringes are then observed along the shorter bandwidth of the joint spectrum (see Fig.\ref{fig:schematic}), rendering the measurement of the slant less stable but that of the spacing more robust. Nevertheless, Fourier analysis retrieves the same values of $\phi_p$, estimated to $(-1.45 \pm 0.2)\cdot 10^5$ fs$^2$ from correlation and $(-1.25 \pm 0.1)\cdot 10^5$ fs$^2$ from reconstruction, consistent with the previous result within the theoretical model. No significant higher-order nonlocal spectral phase was observed.

To confirm the validity of our setup, the experiment was also realized without chirping the pump by bypassing the prism compressor. Because that setup was greatly reducing the amount of spatial chirp \cite{mosley:08}, the idler photon spectral bandwidth was less when the pump was chirped. Consequently, a pulse shaper was used to carve a shorter bandwidth for the unchirped pump measurement. The pulse shaper was optical fiber-coupled, adding additional dispersion to the single photons before the interferometer was open. Consequently, that phase was compensated by the pulse shaper, as described in \cite{PRL:18,PRA:18}. The spectral interference fringes depicted in Fig.\ref{fig:schematic} are clearly no slanted as they are in the chirped pump case, pointing to the absence of phase correlations. Remarkably, the experiment recovered a positive correlated phase of $(7.4 \pm 1) \cdot 10^3$ fs$^2$ for the signal and $(5.5 \pm 2) \cdot 10^3$ fs$^2$ for the idler photon. While small, these values are statistically significant and can be attributed to the pump being slightly chirped and also to correlations caused by the PDC crystal itself (see Methods). This was estimated to $\sim 3500$ fs$^2$ which is in strong agreement with the recovered phase. The reconstructed spectral phase yielded the following coefficients: $(1.4 \pm 0.5)\cdot 10^4$ fs$^2$ for the signal and $(1.2 \pm 0.2)\cdot 10^4$ fs$^2$ for the idler, showing that the pulse shaper was close to compensating the large amount of dispersion caused by the meters of optical fibers leading to the interferometer.

Having measured the uncorrelated spectral phase of each photon as well as the nonlocal phase correlation for a chirped and a unchirped pump, it is now possible to reconstruct the full joint spectral amplitude. The reconstruction, depicted by Fig.\ref{fig:reconstruction} (a,b), shows the amplitude of the probability distribution $\left| f(\omega_1,\omega_2) \right|^2$ of the two-photon state using contours overlaid on a heat-map that shows the phase $\operatorname{Arg} \left( f(\omega_1,\omega_2) \right)$. This is represented for both the unchirped and chirped case. While the amplitude is indeed separable, the phase for a chirped pump is not flat and clearly shows a saddle-like shape. Such a distribution is not possible with separable states, and reveals entanglement from phase only. Note that the amplitude differ between both cases because of the prism compressor in the chirped case that affect in a non-trivial manner the spatial interaction within the SPDC crystal. It still remains largely uncorrelated.
Ultimately, this allows us to describe the entanglement of those states by performing a Schmidt decomposition on the two reconstructed distributions. The first 20 eigenvalues are depicted by Fig.\ref{fig:reconstruction} (c,d) for both cases, where the error bar are built from averaging multiple acquisitions. As a comparison, the modulus of the distributions were also diagonalized and their eigenspectrum is plotted on the same graphs (diamond markers). One can see that the distribution without phase involves only a single mode in both cases ($K = 1.016$ and $1.059$ respectively for the unchirped and chirped distributions). The correlated phase in the chirped case greatly affects that and results in a multimode decomposition as the Schmidt number is increased to $K = 5.0 \pm 0.5$. The singular values, which denote the strength of the entanglement or the squeezing parameter, are distributed over a larger number of singular modes, which are represented in Fig.\ref{fig:reconstruction} (e,f). One can see that for the unchirped case, the first mode is composed of the marginals of the joint spectral amplitude, while the higher order modes are degenerate and negligible. Interestingly, in the chirped case, not only the number of main singular modes is important but also their bandwidth is reduced. This is a consequence of the multimode aspect of the state, since the modes need to be summed to adequately represent the full joint spectral distribution. Remarkably, this shows that only by taking phase into consideration is the high-dimensionality entanglement revealed.

\begin{figure}[t]
	\centering
	\includegraphics[width=\linewidth]{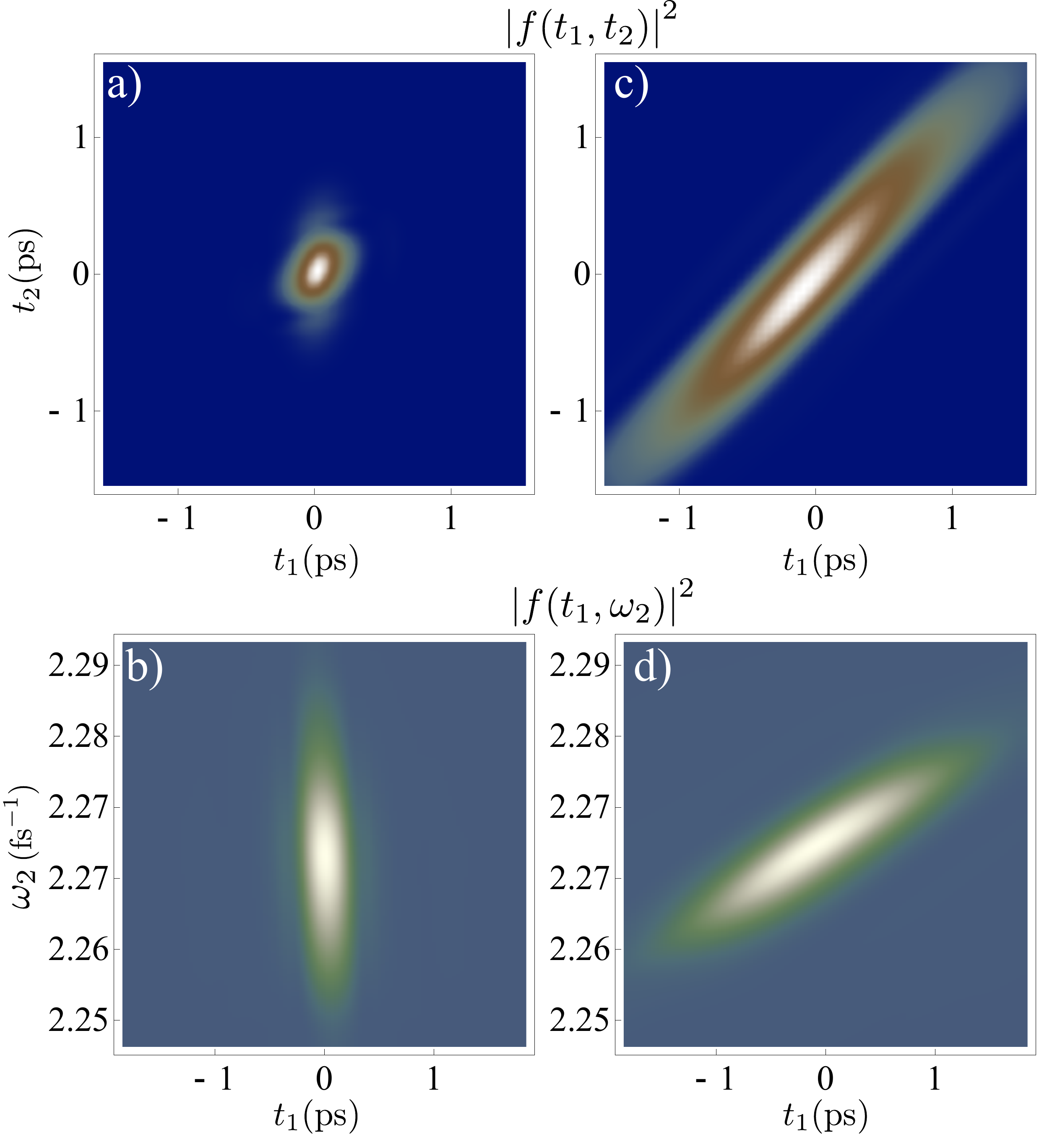}
	\caption{Fourier transform of the joint spectral distributions full (a,c) and partial (b,d). a,b) unchirped; c,d) chirped.}
	\label{fig:entanglement}
\end{figure}

Alternatively, another way of representing entanglement is by Fourier transforming the complex distributions. The resulting distributions are shown in Fig.\ref{fig:entanglement}. The modulus square are shown to conceptualize the outcome of a time-of-flight measurement. As expected, in the unchirped case (a-b), both the joint temporal intensity  and the spectral-temporal intensity are separable. In the chirped case, one can clearly see that the spectral-temporal intensity (d) becomes correlated, since it can be written as $f(t_1 - \phi_p \omega_2)$, which demonstrates time-frequency entanglement. The joint tempoal intensity (c), however, is also correlated. This is direct consequence of the chirp present on both daughter photons, which maps their individual frequencies onto their individual times. Because of the entanglement between each respective photon's time and spectral degree of freedom, this results in this correlated temporal distribution. Note, however, that the sign of the correlation cannot change, and is therefore independent of the sign of the pump chirp. Note also that, although the amount of chirp that was applied to the pump can be considered very large in the ultrafast field, the range over which the correlations in time or in time-frequency are observed is only on the few picoseconds range. This would prove very challenging to measure with standard detectors, while our methods, without utilizing particularly state-of-the-art single photon detectors, takes advantage of interferometry to push this timing information into the spectral domain, allowing to retrieve the effect of the entanglement.

In order to verify that the apparent spectral phase correlations were indeed coming from the dispersion of the pump, we repeated the experiment with the pump bypassing the prism stretcher. In this case, we were able to measure a small but significant positive correlated phase of $\phi_p=9 \pm 5 \times 10^3$ fs$^2$ with the experiment in the configuration where the signal photon is routed into the EOSI. This is consistent with the phase that the pump beam would be expected to accumulate from the various optics up to and including the down conversion crystal. The result from the other configuration was consistent but not statistically significant, since the shorter bandwidth of the heralding photon leads to a less stable reconstruction. These results verify that in this experiment, controllable manipulation of the pump introduced phase correlations in the daughter photons that influenced the degree of entanglement.

To conclude, we have applied the technique of electro-optic spectral shearing interferometry to demonstrate an original, highly flexible and reference-free method to measure the global wave function of a quantum system that exhibits multipartite entanglement across a high-dimensional basis. The unprecedented input acceptance combined with the reliability of the reconstruction illustrates the promise of this technique for understanding broadband quantum interactions. This demonstration, where control of the parameters of the source was shown to have an elusive but powerful effect on the quantum properties of the output state, exemplifies the importance of a full characterisation scheme in understanding the behaviour and tunability of quantum devices. In providing a tool for understanding and hence controlling the behaviour of the elements of a quantum network, this approach has the potential to lead to transformative changes in high-dimensional quantum optical technologies such as TF-encoded crytography, simulation, metrology and communications.

\begin{acknowledgments}
	We are grateful to B. Brecht and M. Karpi\'{n}ski for fruitful discussions and insight on the results. This project has received funding from the European Union's Horizon 2020 research and innovation programme under Grant Agreement No. 665148, the United Kingdom Defense Science and Technology Laboratory (DSTL) under contract No. DSTLX-100092545, and the National Science Foundation under Grant No. 1620822.
\end{acknowledgments}

\bibliography{bibliographyNat}
\onecolumngrid
\pagebreak

\section{Methods}

\subsection{Entanglement measurement}
We use as our measure of entanglement the effective Schmidt rank $K$, defined as
\begin{equation*}
K=\frac{[\sum_i \lambda_i^2 ]^2}{\sum_i \lambda_i^4}
\end{equation*}
For pure states, $K>1$ implies entanglement. If there are $X$ nonzero terms in the Schmidt decomposition, each with equal values of $\lambda_i$, then $K=X$. The effective Schmidt rank therefore has the property that it coincides with the number of modes of entanglement for finite-dimensional systems and  may be interpreted as a generalisation of this to continuous-variable quantum systems such as time-frequency encoded states.

\subsection{Photon pair generation}
An 80 MHz train of femtosecond pulses with duration 100 fs and central wavelength are derived from a Ti:Sapphire oscillator (Spectra-Physics Tsunami). A pick-off from the primary beam is directed onto a fast photodiode to provide a global clock for the experiment. The primary pulse train is frequency-doubled in a bismuth barium borate (BiBO) crystal to generate a second-harmonic beam with central wavelength of 415 nm and average power of 550 mW.  The upconverted beam is then sent into a prism stretcher consisting of 2 equilateral SF11 prisms separated by approximately 30 cm, applying approximately 200,000 fs$^2$ of negative dispersion to the beam. The return beam is picked off with a cut mirror and is focused in a 8-mm-long bulk potassium dihydrogen phosphate (KDP) crystal, cut and oriented for Type-II phase-matched spontaneous parametric down conversion (SPDC) to produce pairs of orthogonally polarized photons with a separable joint spectral intensity.  In the SPDC process, the lowest order phase correlations arise from the dispersion of the pump field. This term has a contribution from the interaction between the pump and the non-linear medium, which is inherent to the SPDC process. It is however possible to control that term by inserting a dispersion compensation scheme on the pump beam, such as a prism or grating compressor\cite{martinez:84}. In our demonstration, this contribution was negligible compared to the quadratic phase deliberately added to the pump by the prism compressor.

\subsection{EOSI} 
The EOSI is implemented in a polarisation Mach-Zehnder design, where the two paths of the interferometer have the same spatial mode but are distinguished by polarization \cite{PRA:18}. This has the advantages of increased phase stability within the interferometer, automatic spatial mode matching, and minimisation of the difference in accumulated spectral phase in each arm. Single photons entering the interferometer are prepared in a superposition of these two polarization modes by a half waveplate oriented with major axes at 45$^\circ$ to the preferred axes of the interferometer. The time delay is controlled by varying the birefringence with a Babinet-Soleil compensator, consisting of two calcite wedges whose position can be varied to alter the thickness of birefringent material in the optical path. Only one of the two polarizations- the one aligned with the extraordinary axis of the EOM- receives spectral shear in the EOM. The two arms of the interferometer are then recombined at a second polarising beam splitter oriented at 45$^\circ$ to the preferred polarisations, with one of the outputs then coupled into a single-photon spectrometer. A comprehensive description of the design, calibration and operation of this device can be found in ref.\cite{PRA:18}.\\

\subsection{Spectrally-resolved coincidence measurement}
The temporally short, spectrally broadband nature of the single-photon wave packets, separated by relatively long and stable intervals, is suitable for a spectrometer based on dispersive time-to-frequency conversion. 
A highly dispersive chirped fibre Bragg grating (CFBG) maps the spectral profile of a single-photon wavepacket onto its time of arrival \cite{davis:17}. The distribution of the delay in arrival time relative to the experiment clock therefore provides us with a spectral measurement of the photons, subject to calibration. Due to the trade-off between quantum efficiency and timing precision in semiconductor avalanche photodiodes, and the importance of precise spectral measurements of the interferograms for the phase reconstruction, different photodetectors were used for the two photons. For the photons emerging from the EOSI, the temporal distribution is determined with resolution better than 50 ps using a pair of free-running, low-timing-jitter single-photon counting modules (Micro Photon Devices, PD-050-CTB-FC) followed by a TDC working in start-stop mode (Fig. 2b). This arrangement provides high spectral resolution of 0.05 nm with a spectral range of approximately 825-835 nm. The spectral measurement on the heralding photon was similarly determined to less than 0.3 nm resolution using a more efficient, but less temporally accurate, photodetector (PerkinElmer, SPCM-AQ4C). A more detailed description of these devices and their calibration can be found in ref.\cite{davis:17}.
\pagebreak
\subsection{Phase reconstruction}
The spectral phase $\phi(\omega_1,\omega_2)$ may be expanded about the central frequencies of the two photons $\omega_{c1}$ and $\omega_{c2}$:
\begin{eqnarray*}
\phi(\omega_1,\omega_2) &=& \phi_{00}+\phi_{10}\delta\omega_1+\phi_{01}\delta\omega_2 \\ &+&\phi_{11}\delta\omega_1\delta\omega_2 +\phi_{20}\delta\omega_1^2+\phi_{02}\delta\omega_2^2+\dots
\end{eqnarray*}
where $\omega_1=\omega_{c1}+\delta\omega_1$, $ \omega_2=\omega_{c2}+\delta\omega_2$, and $ \phi_{00} \equiv \phi(\omega_{c1},\omega_{c2})$, the unphysical global phase.
The phase is reconstructed using a multi-dimensional generalisation of the Fourier analysis methods common to spectral shearing interferometry as described in refs.\cite{walmsley:09} and \cite{PRA:18}. The interference pattern $S_\Omega(\omega_1,\omega_2;\Omega)$ is Fourier transformed with respect to the frequency of the photon sent into the interferometer.  
\begin{eqnarray}
&&\overline{S}(T,\omega_2,\Omega) \equiv \mathcal{FT}_1\{S_\Omega(\omega_1,\omega_2;\Omega)\} \nonumber \\ \nonumber
& = &\frac{1}{2}\overline{S}(T,\omega_2)(1+e^{iT\Omega})\\ \nonumber& + & \mathcal{FT}_1\{|f(\omega_1,\omega_2)f(\omega_1+\Omega,\omega_2)|e^{i \Delta\phi(\omega_1,\omega_2)}\}*\delta(T-\tau)\\ \nonumber &+& \mathcal{FT}_1\{|f(\omega_1,\omega_2)f(\omega_1+\Omega,\omega_2)|e^{-i\Delta\phi(\omega_1,\omega_2)}\}*\delta(T+\tau)
\end{eqnarray}
where $\mathcal{FT}_1$ denotes a Fourier transform with respect to the first variable, $*$ indicates a convolution, $\Delta\phi(\omega_1,\omega_2)=\phi(\omega_1,\omega_2)-\phi(\omega_1+\Omega,\omega_2)$ and $\overline{S}(T,\omega_2) \equiv \mathcal{FT}_1\left[S(\omega_1,\omega_2)\right]$. The latter two terms, which contain the phase information, correspond to peaks in the Fourier domain that are separated from the central peak by the added delay in the interferometer, $\tau$. Since $\tau$ is chosen to be much larger than the temporal duration of the pulse and hence the width of $\overline{S}(T,\omega_2)$, these peaks can be isolated by Fourier-domain filtration. Inverse-Fourier transforming either peak and taking the argument of the resulting term provides the phase differential $\phi(\omega_1+\Omega,\omega_2)-\phi(\omega_1,\omega_2)$. These data may then be treated using the analyses described in ref.\cite{PRA:18} to extract the phase $\phi(\omega_1,\omega_2)$ up to an additive term that is a function of $\omega_2$ only. Nonetheless this is sufficient to determine the nonlocal (correlated) phase. 
The component of the phase that is linear in $\omega_1$, corresponding to the global delay of photon 1, can be extracted and its dependence on $\omega_2$ determined. This coefficient is precisely the lowest-order nonlocal phase $\phi_{11}$. The contribution of this component to $\Delta\phi(\omega_1,\omega_2)$ is $ \phi_{11}\Omega\omega_2$, i.e. the interference fringes along the $\omega_1$ direction have a phase offset that is linear in $\omega_2$. It is this term that creates the diagonal slant in the interference pattern in fig.\ref{fig:schematic}.

The remaining part of the phase, consisting of the terms $\phi_{0n}$, is determined by simply swapping the configuration round and repeating. This leaves only $\phi_{00}$, the physically irrelevant global phase. It is worth noting that this prescripition involves some redundancy, since all terms in the expansion of $\phi(\omega_1,\omega_2)$ that depend on both $\omega_1$ and $\omega_2$ may be determined in either configuration.

\subsection{Error estimation}
A typical acquisition consist of running the TRSPS in coincidences over a few hours at a typical rate of 15 coincidences per second, resulting in coordinates in the $\omega_1,\omega_2$ basis. Those are binned in a 2D-histogram to represent the interferograms depicted by Fig.\ref{fig:schematic}. The length of the subset of coordinate to be binned was set to 5000 points, which roughly translates to 6 minutes of acquisition, resulting in a single interferogram. The phase reconstruction algorithm is then applied to all the individual interferograms from the long acquisition, consisting first of isolating one sideband in the Fourier domain and determining its intensity, or contrast of the interference fringes. This allows to apply a first selection on the data set by determining if the contrast is beyond a certain threshold set to $20\%$. The selected dataset are then analyzed by extracting the phase of the interference fringes over the two dimensions in the spectral domain. That two-dimensionnal phase then undergoes a least square regression method to extract the value of the coefficients for $\phi_2$ and for $\phi_p$.
Since this algorithm is applied to a decent amount of interferograms from the whole data set, statistics are created by averaging the result of every phase extraction, the amplitude of the joint spectrum as well as the results of the singular value decomposition.

\subsection{Noise and stability}
The main source of noise in the experiment that results in the error bars is attributed to phase drift in the interference fringes. Indeed, the whole algorithm is relient on the fact that the absolute phase of the fringes is stable over the data subset that is to be binned. If the phase drifts, the fringes wash away, rendering the reconstruction impossible. This is the reason why a small subset is used for the experiment resulting in relatively low count but in high visibility of the fringes. In theory, the position of the fringes should be set since no path variation can happen in the polarization interferometer. In practice, this phase does drift with time because of different factors in the EOM. Mostly, thermal effect in the waveguide result in a change in the index of refraction of the non-linear medium, which in turn can cause the phase to drift for several radiants over the course of an hours. This effect was mitigated by carefully controlling the temperature variations of the EOM casing by attaching it to a heatsink. Another more serious issue is the accuracy of the phase lock loop that is used to generate the 10 GHz RF-wave within the EOM. Additional phase drift results from an intricate combination of the repetition rate lock of the laser and that of the RF-wave. Our current setup utilizes as best as possible every electronic that is available, but could only be improved with better locking scheme and apparatus. Consequently, the phase drifts, although minimized, were still present to the point were phase information would become lost if one were to analyze an interferogram averaged over a few hours. Note that the setup is still very well stabilized, and the constraint on the acquisition time is also impacted by the poor quantum efficiency of our detectors ($\sim 10\%$).

\begin{figure}[t]
	\centering
	\includegraphics[width=\linewidth]{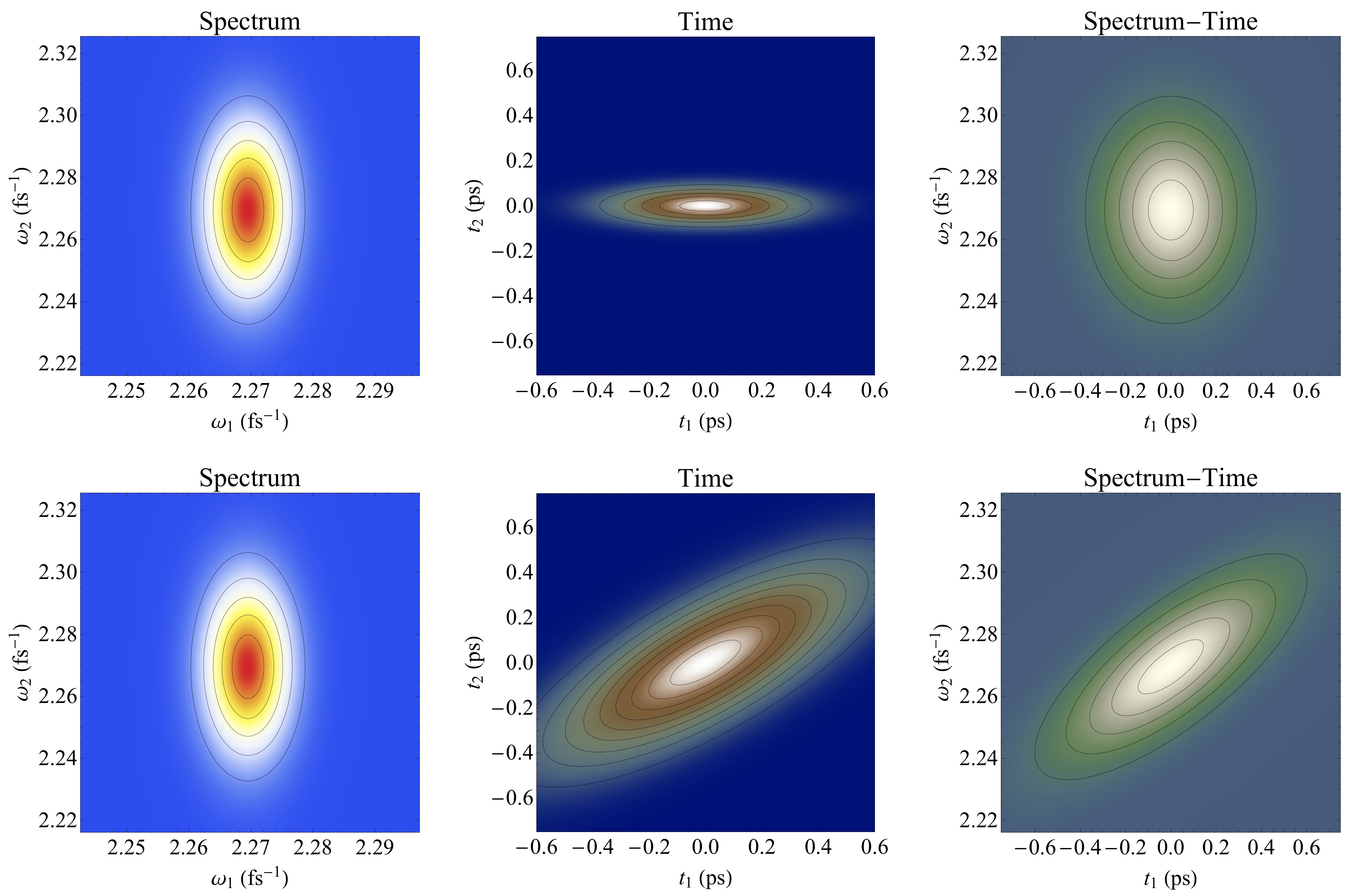}
	\caption{Add section in methods or supplemental explaining the correlations in time.}
	\label{fig:heatmaps}
\end{figure}

\end{document}